\documentclass[letterpaper, 10 pt,conference]{ieeeconf}

\IEEEoverridecommandlockouts
\usepackage[T1]{fontenc}
\usepackage{amsmath,amsfonts,amssymb,mathtools}
\usepackage{xcolor,graphicx}
\usepackage{booktabs,multirow}
\usepackage{cite}
\usepackage{microtype}
\usepackage{paralist}
\usepackage{tikz}
\usepackage{tikz-cd}
\usetikzlibrary{arrows.meta, positioning, calc, shapes, backgrounds}
\tikzcdset{arrow style=tikz, diagrams={>={Stealth}}}
\usepackage[colorlinks=true,
            linkcolor=blue!60!black,
            citecolor=green!45!black,
            urlcolor=blue!70!black]{hyperref}

\newtheorem{theorem}{Theorem}[section]
\newtheorem{proposition}[theorem]{Proposition}

\newtheorem{lemma}[theorem]{Lemma}
\newtheorem{definition}[theorem]{Definition}
\newtheorem{remark}{Remark}[section]
\newtheorem{assumption}{Assumption}[section]

\newcommand{\R}{\mathbb{R}}
\newcommand{\TT}{\mathcal{T}}
\newcommand{\TC}[1]{\mathcal{T}_{\!#1}}
\newcommand{\hcap}{h_{\cap}}
\newcommand{\htcap}{\tilde{h}_{\cap}}

\newcommand{\Safe}{\mathfrak{C}}
\newcommand{\Kl}{\mathbf{Kl}}
\newcommand{\Diff}{\mathbf{Diff}}
\newcommand{\un}{\eta}
\newcommand{\mul}{\mu}
\DeclareMathOperator*{\argmin}{arg\,min}

\title{\LARGE \bf Representing Differential Inclusions as Monads:\\
A Framework for Compositional Controller Synthesis}

\author{
Ahmad Ahmad$^{1}$, Kevin Leahy$^1$
\thanks{$^1$ Department of Robotics Engineering, Worcester Polytechnic Institute, {\tt\small \{aahmad4,kleahy\}@wpi.edu}}
}

\begin{document}

\maketitle

\begin{abstract}

Sequential composition of safety-certified controllers for hybrid and non-smooth robotic systems is well studied piecewise. Yet no existing framework reduces this question to a single algebraic condition. We construct a \emph{trajectory monad} for Filippov-regular differential inclusions. Its Kleisli category treats controllers as morphisms and sequential execution as trajectory concatenation. Control Barrier Function (CBF) certification, a guarantee of forward invariance, corresponds to a sub-monad of this category. Joint safety of composed controllers then follows from one algebraic condition. All controllers must land in a common sub-monad, which is closed under the monad multiplication.

The sub-monad condition recasts recasts existing CBF composition practice categorically. It explains why a smooth intersection of multiple barriers supports safe composition, while enforcing barriers phase by phase does not. For the phase-by-phase case, we derive a closed-form Bernoulli model of the failure rate. We validate it over a randomized sweep of 3{,}000 instances. The same sweep shows that the standard baseline degrades under constraint conflict as the barrier count grows. This baseline enforces all barriers simultaneously as constraints in a single quadratic program (QP). We instantiate the framework on a four-mode piecewise-affine system and on a unicycle reduced-order model over heterogeneous terrain.
\end{abstract}

\section{Introduction}
\label{sec:intro}

Compositional controller design is a fundamental challenge in hybrid robotics, including legged locomotion, manipulation with contacts, and variable-structure control. Given controllers certified to be individually safe, does their sequential execution produce a jointly safe trajectory? Existing approaches answer this piecewise, each under its own composition regime. No unifying account explains why some composition strategies preserve safety certificates while others provably fail.

Several mature lines of work address pieces of this problem. Boolean and smooth CBF compositions~\cite{glotfelter2017nonsmooth, glotfelter2019provably, molnar2023composing, ong2025combinatorial} prove forward invariance for min/max, smooth AND/OR, and nested Boolean combinations of barriers. They are compositional within a fixed operating mode. A separate line of work asks when several CBFs can be enforced jointly without emptying the input-constrained feasible set~\cite{breeden2023compositions, tan2022compatibility}. We observe this QP infeasibility empirically for the simultaneous-constraint baseline in \S\ref{sec:statistical}. Outside the CBF literature, sequential composition has a long history in robotics through funnels built from Lyapunov sublevel sets~\cite{burridge1999sequential}. Sum-of-squares funnel libraries later made this approach robust to disturbances~\cite{majumdar2017funnel}. Assume-guarantee contracts~\cite{saoud2021contracts} provide cascade and feedback composition theorems for continuous-time invariance, with extensions to hybrid systems~\cite{moeller2026hybrid}. Monadic and categorical systems theory~\cite{myers2022categorical, goncharov2018hybrid, goncharov2019duration, jacobs1996coalgebraic} has used Kleisli categories for dynamical-system composition since the work in ~\cite{jacobs1996coalgebraic}. The hybrid monad of ~\cite{goncharov2018hybrid} is the closest predecessor to our construction.

None of these frameworks gives a unified categorical formulation in which CBF certificates lift as sub-monadic structure with provably closed Kleisli composition. The Boolean nonsmooth compositions of~\cite{glotfelter2017nonsmooth} are mode-local and do not address sequential phase transitions. Funnel-based sequential composition~\cite{burridge1999sequential, majumdar2017funnel} checks compatibility funnel by funnel rather than through a single algebraic condition. It also targets Lyapunov sublevel sets rather than CBF-certified safe sets. The input-constrained compatibility conditions of~\cite{breeden2023compositions, tan2022compatibility} guarantee feasibility of the composed QP. Like the Boolean compositions, however, they are stated mode-locally rather than across sequential handoffs. The hybrid monad~\cite{goncharov2018hybrid, goncharov2019duration} handles hybrid trajectory composition but does not embed safety certificates. Assume-guarantee contracts~\cite{saoud2021contracts} give sufficient conditions for composition in a contract-based formalism rather than an algebraic one. Within that framework, the question of why composition fails has no algebraic answer.



\textbf{Contribution.} We develop a trajectory monad that fills this gap. It is a mathematical structure on Filippov-regular differential inclusions. Its Kleisli category is a formal language for composing safety-certified controllers. Its sub-monad structure gives a sufficient condition for compositional safety and a precise account of why naive composition fails. The sub-monad account also has analytical consequences. It identifies the failure mode as a local mismatch of sub-monad membership at each Kleisli handoff. This locality lets us model the failure probability as a sequence of independent trials, one per handoff (\S\ref{sec:statistical}).

Specifically, the contributions of this work are:
\begin{inparaenum} 
  \item Construction of the trajectory monad for Filippov-regular differential inclusions on diffeological spaces, with the monad laws holding exactly (\S\ref{sec:monad}).
  \item Formalization of sequential controller design in the resulting Kleisli category,  which encodes mode switching as Kleisli composition (\S\ref{sec:kleisli}).
   \item Lifting of CBF certificates to sub-monads, a characterization of when Kleisli composition fails, and a sufficient condition for compositional safety. The condition is a common sub-monad, realized by a smooth intersection CBF with an explicit margin (\S\ref{sec:cbf}).
  \item A closed-form Bernoulli model for the naive failure rate, which treats each Kleisli handoff as an independent trial. The sub-monad mismatch structure of \S\ref{sec:cbf} motivates this decomposition. We validate the model over 3{,}000 randomized instances with phase counts $n\in\{2,\dots,6\}$ and barrier counts $k\in\{2,\dots,5\}$. The same sweep shows the standard simultaneous-constraint QP baseline degrading under constraint conflict as $k$ grows (\S\ref{sec:statistical}).
  \item Numerical instantiation on two use cases, a four-mode piecewise-affine (PWA) system and a unicycle reduced-order model on heterogeneous terrain (\S\ref{sec:usecases}).
\end{inparaenum}

\textbf{Related work.} The categorical treatment of open dynamical systems via polynomial functors and lenses is developed in~\cite{myers2022categorical}. Section~2.3 of that work uses Kleisli composition explicitly as a systems-composition language. We specialize to Filippov systems and add safety-certificate lifting, which does not appear there. The hybrid monad of~\cite{goncharov2018hybrid, goncharov2019duration} uses the same Moore-path construction for strictly associative hybrid composition. Its authors use it to define the meaning of loops in hybrid programs. They do not address safety certificates or controller design (see Remark~\ref{rem:duration}). The sheaf-theoretic formulation of~\cite{schultz2020sheaves} offers a complementary account of dynamical-system composition. The smooth Boolean CBF compositions of~\cite{lindemann2019stl, molnar2023composing} and the softmin barrier analysis of~\cite{rabiee2023softmin} are the engineering tools we embed into the monadic framework. We recover their safety guarantee from sub-monad closure (Theorem~\ref{thm:main}).

\section{Preliminaries}
\label{sec:prelim}

\subsection{System Model and Filippov Regularity}
\label{sec:sysmodel}

Let $M$ be a smooth manifold of dimension $d$ representing the state space. Consider the control-affine system
\begin{equation}\label{eq:system}
  \dot{x} = f(x) + B(x)\,u, \quad u \in \mathbf{U} \subset \R^m,
\end{equation}
where $f : M \to TM$ and $B(x) : \R^m \to T_xM$, with $TM$ the tangent bundle of $M$. We assume $f$ and $B$ are locally bounded and measurable, and locally Lipschitz away from finitely many switching surfaces. The input set $\mathbf{U}$ is compact and convex. The dynamics may switch discontinuously, e.g., at contact events or in variable-structure control. The right-hand side of~\eqref{eq:system} then need not admit a classical solution. We instead model such behavior via the Filippov \emph{differential inclusion}~\cite{filippov1988diffeq, cortes2008discontinuous}
\begin{equation}\label{eq:di}
\begin{aligned}
      \dot{x} \in &F(x) := \bigcap_{\rho>0}\,\\&\bigcap_{\mathrm{Leb}(\mathcal{N})=0}\overline{\mathrm{co}}\bigl\{f(y) + B(y)u \mid y \in \mathcal{B}_\rho(x)\setminus\mathcal{N},\ u \in \mathbf{U}\bigr\}.
\end{aligned}
\end{equation}
Here $\overline{\mathrm{co}}\{\cdot\}$ denotes the closed convex hull, and $\mathcal{B}_\rho(x)$ is the ball of radius $\rho$ about $x$ in local coordinates. The inner intersection ranges over all Lebesgue-null sets $\mathcal{N}$. Where $f$ and $B$ are continuous, $F(x)$ reduces to the set of control-reachable velocities $\overline{\mathrm{co}}\{f(x)+B(x)u \mid u \in \mathbf{U}\}$. On a switching surface, it also convexifies the velocities of the adjacent modes. This is what admits absolutely continuous solutions across the switch.

\begin{assumption}[Filippov regularity~\cite{filippov1988diffeq}]
\label{ass:filippov}
  The set-valued map $F : M \rightrightarrows TM$ is upper-semicontinuous with nonempty compact convex values everywhere on $M$.
\end{assumption}

Assumption~\ref{ass:filippov} holds for~\eqref{eq:di} whenever $f$ and $B$ are locally bounded and measurable and $\mathbf{U}$ is compact and convex~\cite{cortes2008discontinuous}. It therefore holds for all PWA systems with box input sets. Under Assumption~\ref{ass:filippov}, Filippov's theorem guarantees that a solution exists on some $[0, r]$ from every $x_0 \in M$. The solution need not be unique.

\subsection{Control Barrier Functions}
\label{sec:cbf_prelim}

Let $h : M \to \R$ be a continuously differentiable function and let $\Safe = \{x \in M : h(x) \geq 0\}$ be the corresponding \emph{safe set}.

\begin{definition}[Control Barrier Function~\cite{ames2019cbf}]
\label{def:cbf}
A function $h$ is a Control Barrier Function (CBF) for system~\eqref{eq:system} if some extended class-$\mathcal{K}$ function $\alpha$ satisfies \[\sup_{u \in \mathbf{U}}\,\bigl[\nabla h(x)^\top \bigl(f(x) + B(x)\,u\bigr)\bigr]\;\geq\; -\alpha\bigl(h(x)\bigr)\] for all $x \in M$. An extended class-$\mathcal{K}$ function is continuous and strictly increasing on all of $\R$, with $\alpha(0) = 0$.
\end{definition}

If $h$ is a CBF, the \emph{CBF-QP safety filter}
\begin{equation}\label{eq:cbfqp}
\begin{aligned}
      u^*(x) = \argmin_{u \in \mathbf{U}}\, \|u - u_{\mathrm{nom}}(x)\|^2\\
   \text{s.t.}\quad
  \nabla h(x)^\top \bigl(f(x) + B(x)\,u\bigr) \geq -\alpha\,h(x)
\end{aligned}
\end{equation}
renders $\Safe$ forward-invariant~\cite{ames2019cbf}. That is, $x(0) \in \Safe$ implies $x(t) \in \Safe$ for all $t \geq 0$. Here $u_{\mathrm{nom}} : M \to \mathbf{U}$ is a nominal (task-level) controller. Throughout, we take the linear choice $\alpha(h) = \alpha h$ with scalar gain $\alpha > 0$, as in~\eqref{eq:cbfqp}. This choice only fixes the implemented constraint. Our results in \S\ref{sec:monad}--\S\ref{sec:cbf} use only forward invariance and hold for any extended class-$\mathcal{K}$ function.

\subsection{Categorical Preliminaries}
\label{sec:catprelim}

We recall the minimal diffeology background used in \S\ref{sec:monad}, together with monads and Kleisli categories. See~\cite{iglesias2013diffeology} and~\cite{myers2022categorical} for the general theories. Throughout, $\circ$ denotes composition of functors, natural transformations, or morphisms in an arbitrary category $\mathcal{C}$, as determined by context.

\begin{definition}[Diffeology]\label{def:diffeology}
A \emph{diffeology} on a set $X$ plays the role for an arbitrary set that a smooth atlas plays for a manifold. It specifies which maps into $X$, from open Euclidean domains, count as smooth. Formally, a diffeology assigns to every open $O \subset \R^p$ (any $p \geq 0$) a set of maps $O \to X$, called \emph{plots}, with three properties. \begin{inparaenum} \item Every constant map $O \to X$ is a plot. \item (Locality) If $O = \bigcup_i O_i$ is an open cover and the restriction of $\lambda : O \to X$ to each $O_i$ is a plot, then $\lambda$ itself is a plot. \item (Smooth reparametrization) If $\lambda : O \to X$ is a plot and $g : O' \to O$ is smooth in the ordinary sense for open $O' \subset \R^{p'}$, then $\lambda \circ g$ is a plot. \end{inparaenum} A set together with a diffeology is a \emph{diffeological space}.
\end{definition}

\begin{definition}[Smooth maps]\label{def:diffsmooth}
A map $q : X \to Y$ between diffeological spaces is \emph{smooth} if $q \circ \lambda$ is a plot of $Y$ for every plot $\lambda$ of $X$. Let $D \subset \R^p$ be a subset that is not open. A map defined on $D$ is smooth if, around every point of $D$, it extends to a smooth map on an open neighborhood in $\R^p$. We write $\Diff$ for the category of diffeological spaces with smooth maps as morphisms.
\end{definition}

Every smooth manifold carries a canonical diffeology whose plots are the smooth maps in the usual sense. This assignment is a full and faithful embedding $\mathbf{Man} \hookrightarrow \Diff$~\cite{iglesias2013diffeology}. We use it to regard $M$ as an object of $\Diff$ without altering its manifold structure. Unlike $\mathbf{Man}$, the category of smooth manifolds, $\Diff$ is complete, cocomplete, and cartesian closed~\cite{iglesias2013diffeology}. Section~\ref{sec:monad} uses these properties.

The following definitions are standard and are stated for a generic monad $(T, \eta, \mu)$ on an arbitrary category $\mathcal{C}$; \S\ref{sec:monad} specializes $\mathcal{C} = \Diff$ and constructs the specific trajectory monad $(\TT, \un, \mul)$.

\begin{definition}[Monad]
\label{def:monad_abstract}
A monad on a category $\mathcal{C}$ is a triple $(T, \eta, \mu)$. Here $T : \mathcal{C} \to \mathcal{C}$ is an endofunctor, $\eta : \mathrm{Id}_{\mathcal{C}}\Rightarrow T$ is a natural transformation called the \emph{unit}, and $\mu : T \circ T \Rightarrow T$ is a natural transformation called the \emph{multiplication}. They satisfy the left unit law $\mu \circ \eta_T = \mathrm{id}_T$, the right unit law $\mu \circ T(\eta) = \mathrm{id}_T$, and the associativity law $\mu \circ \mu_T = \mu \circ T(\mu)$. The functor $\mathrm{Id}_{\mathcal{C}}$ is the identity on $\mathcal{C}$, and $\mathrm{id}_T$ is the identity natural transformation on $T$, with components $(\mathrm{id}_T)_A = \mathrm{id}_{T(A)}$.
\end{definition}

\begin{definition}[Kleisli category]
\label{def:kleisli_abstract}
  The \emph{Kleisli category} $\Kl(T)$ of a monad $(T, \eta, \mu)$ on
  $\mathcal{C}$ has the same objects as $\mathcal{C}$.
  A \emph{Kleisli morphism} $\varphi : X \to Y$ in $\Kl(T)$ is a morphism
  $\varphi : X \to T(Y)$ in $\mathcal{C}$. Two such morphisms
  $\varphi : X \to T(Y)$ and $\chi : Y \to T(Z)$ compose as
  \begin{equation}\label{eq:kleisli_abstract}
    \chi \circ_K \varphi \;:=\; \mu_Z \circ T(\chi) \circ \varphi \;:\; X \to T(Z),
  \end{equation}
  where $\circ_K$ denotes Kleisli composition. 
\end{definition}


\begin{definition}[Sub-monad]
\label{def:submonad_abstract}
  A sub-monad of $(T, \eta, \mu)$ is a monad $(T', \eta', \mu')$ together with a monad morphism $\iota : T' \hookrightarrow T$, which is a natural transformation commuting with units and multiplications. We say $T'$ is a sub-monad of $T$ and write $T' \subset T$.
\end{definition}

\subsection{Running Examples}
\label{sec:examples}

We consider two systems as use cases throughout the paper. Both satisfy Assumption~\ref{ass:filippov}, so the trajectory monad construction of \S\ref{sec:monad} applies to each.

\textbf{Use Case I (four-mode PWA system).} Let $x = (x_1, x_2)^\top \in \R^2$ be the state. The system is $\dot{x} = A_{\ell(x)} x + c_{\ell(x)} + u$ with $u \in \mathbf{U} = [-1,1]^2$. The mode index $\ell(x) \in \{1,2,3,4\}$ is the quadrant of $x$~\cite{belta2017formal}. Each mode combines a contraction with a rotation, and the rotation direction alternates between adjacent quadrants. We use $A_1 = A_3 = \begin{bmatrix} -0.4 & 1.5 \\ -1.5 & -0.4 \end{bmatrix}$ and $A_2 = A_4 = A_1^\top$. The offsets are $c_1 = (0.2, 0.2)^\top$, $c_2 = (-0.2, 0.2)^\top$, $c_3 = -c_1$, and $c_4 = -c_2$, so each offset points away from the origin into its own quadrant. This corresponds to $f(x) = A_{\ell(x)} x + c_{\ell(x)}$ and $B(x) = I_2$ in~\eqref{eq:system}. In the interior of each quadrant, the map~\eqref{eq:di} is $F(x) = A_{\ell(x)} x + c_{\ell(x)} + [-1,1]^2$. On the coordinate axes, it is the convex hull of the adjacent modes' velocity sets. In both cases $F(x)$ is a compact convex polytope.

\textbf{Use Case II (modal unicycle reduced-order model).} Let $\xi = (p_x, p_y, \psi)^\top \in M := \R^2 \times S^1$, where $(p_x, p_y)$ is the planar position and $\psi$ is the heading angle. The control input is $u = (v, \omega)^\top$, with forward speed $v$ and yaw rate $\omega$. The unicycle reduced-order model (ROM)~\cite{cohen2024rom} has input matrix $G(\xi) = \begin{bmatrix}\cos\psi & 0\\ \sin\psi & 0\\ 0 & 1\end{bmatrix}$. We add terrain modes that make the plant a genuine Filippov system. The mode $\ell(\xi) = \arg\min_i h_i(\xi) \in \{1,2,3\}$ indexes the most binding of the three barriers of \S\ref{sec:usecases}, which represent flat ground, stepping stones, and a slope edge. The dynamics are $\dot{\xi} = f_{\ell(\xi)} + G(\xi)\,u$ with $u \in \mathbf{U}_{\ell(\xi)} = [0, v_{\max,\ell}] \times [-\omega_{\max,\ell}, \omega_{\max,\ell}]$. The limits $(v_{\max,\ell}, \omega_{\max,\ell})$ are $(0.45, 1.50)$, $(0.25, 0.90)$, and $(0.35, 1.10)$ for $\ell = 1, 2, 3$, in m/s and rad/s. Only the stepping stones carry drift, $f_2 = (0.015, 0, 0)^\top$\,m/s, while $f_1 = f_3 = 0$. This corresponds to $f = f_{\ell(\xi)}$ and $B = G$ in~\eqref{eq:system}.

The modes switch on the surfaces $\{h_i = h_j\}$, where both the input set and the drift jump. The drift $f_2$ lies outside the span of $G(\xi)$ unless the robot heads along the $p_x$-axis. The velocity sets of mode 2 and its neighbors are then disjoint, so~\eqref{eq:di} convexifies across a genuine discontinuity. Assumption~\ref{ass:filippov} holds because the drift is bounded and each $\mathbf{U}_\ell$ is compact and convex. Cohen et al.~\cite{cohen2024rom} show that the unicycle ROM captures the safety-relevant navigation behavior of an 18-DOF quadruped.

\section{The Trajectory Monad}
\label{sec:monad}



Definition~\ref{def:kleisli_abstract} presupposes a monad on the systems of interest. Before asking whether CBF certificates are preserved under sequential composition, we construct one. We write $(\TT, \un, \mul)$ for this specific trajectory monad, to distinguish it from the generic monad $(T, \eta, \mu)$ of Definitions~\ref{def:monad_abstract}--\ref{def:submonad_abstract}. The endofunctor $\TT$ sends each state to the trajectories it can generate. The unit $\un$ regards a state as the trivial trajectory at that state. The multiplication $\mul$ flattens a trajectory of trajectories. It follows the initial points of the inner trajectories and then appends the last inner trajectory.

We construct $\TT(M)$ in $\Diff$, the category of diffeological spaces~\cite{iglesias2013diffeology}. The space of all admissible trajectories of~\eqref{eq:di} is infinite-dimensional and not itself a manifold. The category $\Diff$, by contrast, is complete, cocomplete, and cartesian closed, and it contains every smooth manifold as a full subcategory~\cite{iglesias2013diffeology}. The state space $M$ therefore remains the finite-dimensional manifold of \S\ref{sec:sysmodel}. Only $\TT(M)$ needs the added generality.

The one genuine difficulty is associativity of concatenation. Suppose trajectories are parametrized on a fixed interval such as $[0,1]$. Concatenating two of them then requires rescaling each by a factor of two. Concatenating a third depends on which pair is rescaled first. Associativity then holds only up to a reparametrizing homeomorphism, not as literal equality of functions. \emph{Moore trajectories} keep their true duration instead of being rescaled onto a fixed interval. They give strict, on-the-nose associativity (\S\ref{sec:moore}).

Throughout, $\TT$ denotes this specific trajectory monad and $\Kl(\TT)$ its Kleisli category. Strictly speaking, $\TT$ acts on pairs $(M, F_M)$, since $\TT(M)$ contains only trajectories admissible for $F_M$. It acts on smooth maps $q : M \to N$ that send admissible trajectories of $F_M$ to admissible trajectories of $F_N$~\cite{tabuada2009verification}. We suppress $F$ from the notation.

\subsection{Moore Trajectories and Strict Associativity}
\label{sec:moore}

\begin{definition}[Moore trajectory space]\label{def:moore}
A \emph{Moore trajectory of duration $r \geq 0$} in $M$ is a pair $(\gamma, r)$, where $\gamma : [0,r] \to M$ is absolutely continuous. That is, $\gamma$ is differentiable almost everywhere, its derivative $\dot\gamma$ lies in $L^1([0,r])$, the space of Lebesgue-integrable functions on $[0,r]$, and $\gamma(t) = \gamma(0) + \int_0^t \dot\gamma(s)\,ds$. The trajectory monad object is the disjoint union over all durations
  \begin{equation}\label{eq:TM}
    \TT(M) = \bigsqcup_{r \geq 0}
    \bigl\{(\gamma, r) \mid \gamma : [0,r] \to M,\;
    \dot\gamma(t) \in F(\gamma(t)) \text{ a.e.}\bigr\}.
  \end{equation}
The endpoint map is $\mathrm{end}(\gamma, r) = \gamma(r)$. We equip $\TT(M)$ with the \emph{functional diffeology}. A map $\lambda : O \to \TT(M)$, with $O \subset \R^p$ open and $\lambda(z) = (\gamma_z, r(z))$, is a plot iff the duration map $z \mapsto r(z)$ is smooth and the evaluation map $(z, t) \mapsto \gamma_z(t)$ is smooth in the sense of Definition~\ref{def:diffsmooth}.
\end{definition}

Consider two Moore trajectories $(\gamma_1, r_1)$ and $(\gamma_2, r_2)$ with $\gamma_1(r_1) = \gamma_2(0)$. Their concatenation is defined by
\[
  (\gamma_1 \cdot \gamma_2)(t) :=
  \begin{cases}
    \gamma_1(t) & 0 \leq t < r_1, \\
    \gamma_2(t - r_1) & r_1 \leq t \leq r_1 + r_2.
  \end{cases}
\]
The concatenation has duration $|\gamma_1 \cdot \gamma_2| = r_1 + r_2$. Durations are real numbers, so $(\gamma_1 \cdot \gamma_2) \cdot \gamma_3 = \gamma_1 \cdot (\gamma_2 \cdot \gamma_3)$ holds \emph{strictly as functions}. Both sides have duration $r_1 + r_2 + r_3$ and agree on every sub-interval.

\begin{remark}[Relation to the hybrid monad]\label{rem:duration}
The hybrid monad of~\cite{goncharov2018hybrid,goncharov2019duration} pairs a duration with a path in a similar way, and our multiplication follows theirs. They use it for strictly associative iteration semantics of hybrid computation, in a deterministic setting. Our construction transfers it to Filippov-admissible trajectories on diffeological spaces. It also lifts safety certificates as sub-monadic structure (\S\ref{sec:cbf}), which neither prior work does.
\end{remark}

\subsection{Unit and Multiplication}
\label{sec:unitmult}

Definition~\ref{def:monad_abstract} requires a unit $\eta$ and a multiplication $\mu$ in addition to the endofunctor. Having constructed $\TT$, we define these two pieces next.

\begin{definition}[Unit]\label{def:unit}
The unit $\un_M : M \to \TT(M)$ maps each state to the duration-zero constant trajectory at that state,
  \[
    \un_M(x_0) = (\mathrm{const}_{x_0},\, 0),
  \]
where $\mathrm{const}_{x_0} : [0,0] \to M$ is the trivial path.
\end{definition}

\begin{definition}[Multiplication]\label{def:mul}
Let $\mathrm{head}(\gamma, r) := \gamma(0)$ denote the initial point of a Moore trajectory. An element of $\TT(\TT(M))$ is an \emph{outer} Moore trajectory $\Gamma : [0,R] \to \TT(M)$ whose head path $\theta \mapsto \mathrm{head}(\Gamma(\theta))$ is admissible for $F$. The multiplication follows the head path up to time $R$ and then appends the last inner trajectory, \[\mul_M(\Gamma) := \bigl(\mathrm{head}\circ\Gamma,\, R\bigr) \cdot \Gamma(R).\] The concatenation is well defined because the head path ends at $\mathrm{head}(\Gamma(R))$. This is the multiplication of the hybrid monad of~\cite{goncharov2018hybrid}, transferred to Filippov-admissible trajectories.
\end{definition}

\subsection{Monad Laws}
\label{sec:monadlaws}

Having fixed $\TT$, $\un$, and $\mul$, it remains to check the monad laws, including naturality of $\un$ and $\mul$. For set-valued $F$, naturality holds only under a regularity condition on $F$. Filippov regularity (Assumption~\ref{ass:filippov}) supplies this condition.

For a map $q$ as above, define $\TT(q)(\gamma, r) = (q \circ \gamma, r)$. By the chain rule for absolutely continuous curves, $\TT(q)$ maps admissible trajectories of $F_M$ to admissible trajectories of $F_N$.

Naturality of $\un$ and $\mul$ means that such a map $q$ commutes with the unit (``doing nothing'') and with concatenation, as in~\eqref{eq:nat_unit}. It certifies that the monad structure does not depend on the choice of coordinates.
\begin{equation}\label{eq:nat_unit}
\begin{tikzcd}[column sep=3.5em, row sep=2.5em,
                baseline=(current bounding box.center)]
  M
    \ar[r, "q"]
    \ar[d, "\un_M"']
  &
  N
    \ar[d, "\un_N"]
  \\
  \TT(M)
    \ar[r, "\TT(q)"']
  &
  \TT(N)
\end{tikzcd}
\qquad
\begin{tikzcd}[column sep=3.5em, row sep=2.5em,
                baseline=(current bounding box.center)]
  \TT^2(M)
    \ar[r, "\TT^2(q)"]
    \ar[d, "\mul_M"']
  &
  \TT^2(N)
    \ar[d, "\mul_N"]
  \\
  \TT(M)
    \ar[r, "\TT(q)"']
  &
  \TT(N)
\end{tikzcd}
\end{equation}

\begin{theorem}[Monad laws]\label{thm:monad}
Under Assumption~\ref{ass:filippov}, $(\TT, \un, \mul)$ is a monad on $\Diff$, and the following diagrams commute on the nose.
  \begin{equation}\label{eq:monad_laws}
  \begin{tikzcd}[column sep=3em, row sep=2.3em,
                  baseline=(current bounding box.center)]
    \TT^3(M)
      \ar[r, "\TT(\mul)"]
      \ar[d, "\mul_\TT"']
    &
    \TT^2(M)
      \ar[d, "\mul"]
    \\
    \TT^2(M)
      \ar[r, "\mul"']
    &
    \TT(M)
  \end{tikzcd}
  \qquad
  \begin{tikzcd}[column sep=2.8em, row sep=2.3em,
                  baseline=(current bounding box.center)]
    \TT(M)
      \ar[r, "\un_\TT"]
      \ar[dr, "\mathrm{id}"']
    &
    \TT^2(M)
      \ar[d, "\mul"]
    \\
    &
    \TT(M)
  \end{tikzcd}
  \end{equation}
\end{theorem}

\begin{proof}[Sketch]
\emph{Unit laws.} For $\Gamma = \un_{\TT(M)}(\gamma, r)$, the head path has duration zero and $\mul_M$ returns $(\gamma, r)$. For $\Gamma = \TT(\un_M)(\gamma, r)$, the head path is $\gamma$ itself and the appended trajectory has duration zero. \emph{Associativity.} For $\Gamma \in \TT^3(M)$, both paths around the left square in~\eqref{eq:monad_laws} produce the same three segments in the same order. These are the heads of the heads of $\Gamma$, the head path of $\Gamma(R)$, and the last inner trajectory of $\Gamma(R)$. Moore concatenation is strictly associative, so the two results agree as functions. \emph{Naturality.} The map $\TT(q)$ acts pointwise on paths, preserves durations, and commutes with $\mathrm{head}$. It therefore commutes with $\un$ and $\mul$, which is exactly~\eqref{eq:nat_unit}. \hfill$\square$
\end{proof}

\section{The Kleisli Category as Controller Language}
\label{sec:kleisli}
We now use the Kleisli category of $(\TT, \un, \mul)$ to model controllers as morphisms and sequential execution as their composition. This gives the language in which \S\ref{sec:cbf} states and proves compositional safety.

Apply Definition~\ref{def:kleisli_abstract} to the trajectory monad $(\TT, \un, \mul)$, with $M, N, Z$ in the role of the generic objects $X, Y, Z$. A \emph{Kleisli morphism} $\varphi : M \to \TT(N)$ assigns to each initial state one admissible Moore trajectory. We call a Kleisli morphism $\varphi : M \to \TT(M)$ a \emph{controller} when it is rooted, i.e., $\mathrm{head}(\varphi(x)) = x$, so that its trajectory starts where it is invoked. Every closed-loop system generates a rooted morphism, and the unit $\un_M$ is rooted. For controllers $\varphi$ and $\chi$, Definition~\ref{def:mul} gives \[(\chi \circ_K \varphi)(x) = \varphi(x) \cdot \chi\bigl(\mathrm{end}(\varphi(x))\bigr).\] Indeed, since $\chi$ is rooted, the head path of $\TT(\chi)(\varphi(x))$ is $\varphi(x)$ itself, and the appended trajectory is $\chi$ run from the endpoint. Kleisli composition is therefore sequential execution. We run $\varphi$, then run $\chi$ from where $\varphi$ stopped. The composite is again rooted, and by Theorem~\ref{thm:monad}, $\circ_K$ is associative on the nose.

Diagram~\eqref{eq:kleisli_diag} shows Kleisli composition in $\Kl(\TT)$.
\begin{equation}\label{eq:kleisli_diag}
\begin{tikzcd}[column sep=4em, row sep=2.5em,
                baseline=(current bounding box.center)]
  M
    \ar[r, "\varphi"]
    \ar[drr, bend right=18, "\chi \circ_K \varphi"']
  &
  \TT(N)
    \ar[r, "\TT(\chi)"]
  &
  \TT^2(Z)
    \ar[d, "\mul_Z"]
  \\
  & &
  \TT(Z)
\end{tikzcd}
\end{equation}

\begin{remark}[Prior uses of Kleisli composition for systems]
\label{rem:kleisli_prior}
Kleisli categories for dynamical systems appear in~\cite{jacobs1996coalgebraic} and in Section~2.3 of~\cite{myers2022categorical}. What is new here is that the morphisms carry CBF safety certificates. A morphism $\varphi : M \to \TC{\Safe}(N)$, with $\TC{\Safe}$ the CBF sub-monad of \S\ref{sec:cbf}, is a controller with a proven invariance guarantee. The sub-monad structure of \S\ref{sec:cbf} is what makes these certificates composable.
\end{remark}

\section{CBF Sub-Monads and Compositional Safety}
\label{sec:cbf}
Kleisli composition models sequential execution, but it does not by itself preserve safety certificates. We now identify which Kleisli morphisms carry a safety guarantee. We then ask when composing two such morphisms preserves it.

\subsection{Lifting CBF Certificates to Sub-Monads}

Let $h : M \to \R$ be a CBF with safe set $\Safe = \{x : h(x) \geq 0\}$
as in \S\ref{sec:cbf_prelim}.

\begin{definition}[CBF sub-monad]\label{def:cbfsubmonad}
The \emph{CBF sub-monad} associated to $h$ is \[\TC{\Safe}(M) = \{(\gamma,r) \in \TT(M) \mid h(\gamma(t)) \geq 0,\ \forall t \in [0,r]\}.\] It collects the admissible Moore trajectories that never leave $\Safe$. Its unit is $\un_M$ restricted to $\Safe$.
\end{definition}

\begin{proposition}[Sub-monad closure]\label{prop:submonad}
The unit and multiplication of $\TT$ preserve $\TC{\Safe}$. That is, $\un_M(\Safe) \subseteq \TC{\Safe}(M)$, and $\mul_M$ maps $\TC{\Safe}^2(M)$ into $\TC{\Safe}(M)$. Hence $\TC{\Safe}$ is a sub-monad of $\TT$, and the inclusion $\iota : \TC{\Safe} \hookrightarrow \TT$ is a monad morphism. Diagram~\eqref{eq:submonad_diag} expresses the compatibility with multiplication.

  \begin{equation}\label{eq:submonad_diag}
  \begin{tikzcd}[column sep=3.5em, row sep=2.5em,
                  baseline=(current bounding box.center)]
    \TC{\Safe}^2(M)
      \ar[r, "\mul^\Safe_M"]
      \ar[d, "\iota \circ \TT(\iota)"']
    &
    \TC{\Safe}(M)
      \ar[d, "\iota"]
    \\
    \TT^2(M)
      \ar[r, "\mul_M"']
    &
    \TT(M)
  \end{tikzcd}
  \end{equation}
\end{proposition}

\begin{proof}
For $x_0 \in \Safe$, the unit $\un_M(x_0) = (\mathrm{const}_{x_0}, 0)$ never leaves $\Safe$. Now let $\Gamma \in \TC{\Safe}^2(M)$. Each $\Gamma(\theta)$ lies in $\TC{\Safe}(M)$, so its head lies in $\Safe$ and the head path never leaves $\Safe$. The appended trajectory $\Gamma(R)$ also stays in $\Safe$. By the piecewise definition of concatenation, $\mul_M(\Gamma)$ stays in $\Safe$ throughout. Diagram~\eqref{eq:submonad_diag} commutes by construction. \hfill$\square$
\end{proof}

Closure uses only the definition of $\TC{\Safe}$, not the CBF property. The CBF property enters in Theorem~\ref{thm:main}. There it guarantees that the CBF-QP controller~\eqref{eq:cbfqp} generates a trajectory in $\TC{\Safe}(M)$ from every $x_0 \in \Safe$. This is the viability of $\Safe$ under~\eqref{eq:di} in the sense of~\cite{aubin1991viability}.

\subsection{Failure of Naive Composition}

\begin{proposition}[Naive composition fails]\label{prop:naive_fails}
  Let $\varphi : M \to \TC{\Safe_1}(M)$ and $\chi : M \to \TC{\Safe_2}(M)$ be controllers certified for different safe sets $\Safe_1 \neq \Safe_2$. The Kleisli composite $(\chi \circ_K \varphi)(x)$ need not lie in $\TC{\Safe_1 \cap \Safe_2}(M)$. Diagram~\eqref{eq:naive_diag} shows why. The two controllers reach $\TT(M)$ through two \emph{different} inclusions, and no common sub-monad need contain the image.

  \begin{equation}\label{eq:naive_diag}
  \begin{tikzcd}[column sep=3.5em, row sep=2em,
                  baseline=(current bounding box.center)]
    &
    \TC{\Safe_1}
      \ar[dr, hook, "\iota_1"]
    &
    \\[-8pt]
    M
      \ar[ur, "\varphi"]
      \ar[dr, "\chi"']
    &
    &
    \TT(M)
    \\[-8pt]
    &
    \TC{\Safe_2}
      \ar[ur, hook, "\iota_2"']
    &
  \end{tikzcd}
  \end{equation}
\end{proposition}

\begin{proof}
Take $M = \R$ with $\dot{x} = u$ and $\mathbf{U} = [-1,1]$. Let $h_1(x) = x$ and $h_2(x) = 1 - x$. Let $\varphi$ be the CBF-QP~\eqref{eq:cbfqp} for $h_1$ with $u_{\mathrm{nom}} \equiv 1$ and duration $2$. From $x_0 = 0.5 \in \Safe_1 \cap \Safe_2$, the constraint $u \geq -\alpha x$ never binds. The trajectory $\varphi(x_0)$ therefore reaches $x = 2.5 \notin \Safe_2$ while remaining in $\Safe_1$. Any composite $\chi \circ_K \varphi$ contains this segment and leaves $\Safe_1 \cap \Safe_2$. \hfill$\square$
\end{proof}

\subsection{Compositional Safety via a Shared Sub-Monad}

Proposition~\ref{prop:naive_fails} is avoided if all phase controllers map into the same sub-monad. We achieve this with a smooth intersection of the individual CBFs, a known construction~\cite{lindemann2019stl, rabiee2023softmin, molnar2023composing}. We add an explicit margin to it.

Let $h_1, \ldots, h_k$ be CBFs with safe sets $\Safe_1, \ldots, \Safe_k \subset M$. For a smoothing parameter $\varepsilon > 0$, the smooth minimum (softmin) of the barriers is $\hcap(x) := -\varepsilon\ln \sum_{i=1}^k e^{-h_i(x)/\varepsilon}$. The shifted intersection CBF is $\htcap(x) = \hcap(x) - \delta$, with margin $\delta := \varepsilon\ln k$.


\begin{remark}[The margin and its role]\label{rem:shift}
The classical log-sum-exp bound~\cite{rabiee2023softmin} gives $\min_i h_i(x) - \varepsilon\ln k \leq \hcap(x) \leq \min_i h_i(x)$. The upper bound yields $\{\hcap \geq 0\} \subseteq \Safe_1 \cap \cdots \cap \Safe_k$. In continuous time, every controller certified by $\hcap$ thus lands in a sub-monad of $\TC{\bigcap_i \Safe_i}$. The margin strengthens this to $\min_i h_i(x) \geq \delta$ on $\{\htcap \geq 0\}$. By the lower bound, $\min_i h_i(x) \leq 2\varepsilon\ln k$ on the boundary $\{\htcap = 0\}$, so the choice $\delta = \varepsilon\ln k$ at most doubles the inherent softmin gap. The margin matters in implementation. Our controllers are sampled-data with zero-order hold, and the continuous-time CBF guarantee does not transfer exactly to this setting~\cite{breeden2022sampled}. The margin $\delta$ absorbs this inter-sample drift at our step sizes. Deriving $\delta$ from a sampled-data bound as in~\cite{breeden2022sampled} would make this rigorous.
\end{remark}

\begin{theorem}[Compositional safety]\label{thm:main}
Let $\varphi_1, \ldots, \varphi_n : M \to \TT(M)$ be CBF-QP controllers~\eqref{eq:cbfqp} designed with the shifted intersection CBF $\htcap$. Their nominal controllers may differ. Assume $\htcap$ is a CBF for~\eqref{eq:system} (Definition~\ref{def:cbf}). Then every $x_0$ with $\htcap(x_0) \geq 0$ satisfies \[(\varphi_n \circ_K \cdots \circ_K \varphi_1)(x_0) \in \TC{\,\Safe_1 \cap \cdots \cap \Safe_k}(M).\] Diagram~\eqref{eq:fix_chain} shows the composition chain.

  \begin{equation}\label{eq:fix_chain}
  \begin{tikzcd}[column sep=2.2em, row sep=2em,
                  baseline=(current bounding box.center)]
    M
      \ar[r, "\varphi_1"]
    &
    \TC{\{\htcap \geq 0\}}
      \ar[r, "\varphi_2 \circ_K"]
    &
    \cdots
      \ar[r, "\varphi_n \circ_K"]
    &
    \TC{\{\htcap \geq 0\}}
      \ar[r, hook, "\iota"]
    &
    \TC{\bigcap_i \Safe_i}
  \end{tikzcd}
  \end{equation}
\end{theorem}
\begin{proof}
Since $\htcap$ is a CBF, each $\varphi_j$ keeps trajectories inside $\{\htcap \geq 0\}$ when started there~\cite{ames2019cbf}. Each such trajectory is an element of $\TC{\{\htcap \geq 0\}}(M)$. Every handoff state is the endpoint of such a trajectory, so it also lies in $\{\htcap \geq 0\}$. Each phase therefore starts where its certificate applies. By Proposition~\ref{prop:submonad}, $\TC{\{\htcap \geq 0\}}$ is closed under $\mul_M$. The Kleisli composite thus lands in $\TC{\{\htcap \geq 0\}}(M)$, as in diagram~\eqref{eq:fix_chain}. Finally, Remark~\ref{rem:shift} gives $\{\htcap \geq 0\} \subseteq \bigcap_i \Safe_i$, and the inclusion $\TC{\{\htcap \geq 0\}} \hookrightarrow \TC{\bigcap_i \Safe_i}$ completes the proof. \hfill$\square$

\end{proof}

Theorem~\ref{thm:main} gives a sufficient condition. Controllers in different sub-monads may still compose safely from particular initial states, as the nonzero Naive safety rates in \S\ref{sec:statistical} show. The hypothesis that $\htcap$ is a CBF is not automatic under input bounds~\cite{rabiee2023softmin, breeden2023compositions}. We monitor it empirically through QP feasibility.

\begin{remark}[Automatic activation]\label{rem:autoact}
At state $x$, the CBF-QP constraint imposed by $\htcap$ is $\sum_{i=1}^k w_i(x)\,\nabla h_i(x)^\top \bigl(f(x) + B(x)\,u\bigr) \geq -\alpha\,\htcap(x)$. The softmin weights are $w_i(x) = e^{-h_i(x)/\varepsilon}/\!\sum_j e^{-h_j(x)/\varepsilon}$. Let $i^*(x) \in \arg\min_i h_i(x)$ index the most binding barrier at $x$. When $h_j(x) \gg 0$ for all $j \neq i^*$, the weight $w_{i^*}(x)$ approaches $1$. The constraint then concentrates on the most binding barrier without any scheduling logic. This lazy activation, observed in~\cite{molnar2023composing, rabiee2023softmin}, is a property of the softmin gradient. The sub-monad view adds that it needs no coordination across phases, since every phase controller enforces the same constraint.
\end{remark}

\section{Statistical Validation}
\label{sec:statistical}

We compare three strategies. \emph{Naive} has phase $j$ enforce only $h_{\sigma(j)}$, with $\sigma(j) := ((j-1) \bmod k)+1$, so successive controllers land in different sub-monads (Proposition~\ref{prop:naive_fails}). \emph{Global} enforces all $k$ barriers as hard constraints in one QP~\cite{ames2019cbf, breeden2023compositions} and lands in $\TC{\bigcap_i \Safe_i}$ whenever that QP is feasible. \emph{Monadic} enforces the single constraint $\htcap \geq 0$ and lands in $\TC{\{\htcap \geq 0\}}$ (Theorem~\ref{thm:main}). Inputs are two-dimensional, so we solve each QP exactly by enumerating its candidate minimizers. If none is feasible, the saturated nominal input is applied, which is what exposes Global to violations. Proposition~\ref{prop:naive_fails} shows that Naive composition can fail, and Theorem~\ref{thm:main} shows that Monadic composition cannot. Neither result says how often Naive fails. We now quantify this failure probability (\S\ref{sec:bernoulli}) and validate all three strategies over a randomized sweep (\S\ref{sec:sweep}).

\subsection{Bernoulli Failure Model}
\label{sec:bernoulli}

Under Naive, phase $j$ enforces only $h_{\sigma(j)}$. By forward invariance, the trajectory stays in $\Safe_{\sigma(j)}$ throughout phase $j$. Nothing constrains the other $k-1$ barriers. At time $\tau_j$, the handoff passes a state certified only for $\Safe_{\sigma(j)}$ to a controller certified only for $\Safe_{\sigma(j+1)}$. Proposition~\ref{prop:naive_fails} identifies these handoffs as the source of the sub-monad mismatch. This motivates modeling each handoff as a separate trial.

\begin{lemma}[Bernoulli failure model]\label{lem:bernoulli}
For $j = 1,\ldots,n-1$, let $E_j$ be the event that the Naive trajectory leaves $\bigcap_i \Safe_i$ during phase $j+1$, i.e., after handoff $j$. Assume that phase $1$, which starts from $x_0 \in \bigcap_i \Safe_i$, is jointly safe. The Naive trajectory is then jointly safe iff $\bigcap_{j=1}^{n-1} E_j^{c}$ occurs. Suppose $E_1,\ldots,E_{n-1}$ are mutually independent with common probability $p_0 = P(E_j) \in [0,1)$. Then
\begin{equation}\label{eq:bernoulli}
P_{\mathrm{safe}}^{\mathrm{Naive}}(n) \;=\; (1 - p_0)^{n-1}.
\end{equation}
This has two consequences. \begin{inparaenum} \item If $p_0 > 0$, then $P_{\mathrm{safe}}^{\mathrm{Naive}}(n) \to 0$ geometrically in $n$. \item The index of the first failing handoff has mean $1/p_0$. The expected number of safe handoffs before the first failure is therefore $(1-p_0)/p_0$. \end{inparaenum}
\end{lemma}

\begin{proof}
By independence, $P\bigl(\bigcap_{j=1}^{n-1} E_j^{c}\bigr) = \prod_{j=1}^{n-1} (1 - p_0) = (1-p_0)^{n-1}$. This gives~\eqref{eq:bernoulli} and claim~1. The index of the first failing handoff is geometric with success probability $p_0$, so its mean is $1/p_0$. Excluding the failing handoff itself gives claim~2. \hfill$\square$
\end{proof}

\begin{remark}[Geometric interpretation of $p_0$]\label{rem:p0geom}

Independence and a common $p_0$ are modeling assumptions that we impose and then test empirically. They are plausible when each phase nearly converges to its CLF target before the handoff, which decorrelates successive handoff states. They are also plausible when barriers are drawn i.i.d., as in the sweep below. We test them empirically in \S\ref{sec:sweep}, where any residual phase-1 failures are absorbed into the fitted $p_0$. Suppose further that the handoff state is roughly uniform on $\Safe_{\sigma(j)}$. A first-order estimate is then $p_0 \approx 1 - \mathrm{Vol}(\bigcap_i \Safe_i)/\mathrm{Vol}(\Safe_{\sigma(j)})$. This is the fraction of the active safe set that lies outside the joint safe set. Adding a barrier can only shrink $\bigcap_i \Safe_i$, so the estimate is nondecreasing in $k$. This agrees with the fits reported below.
\end{remark}

\subsection{Randomized Sweep}
\label{sec:sweep}
\textbf{Setup.} We sweep $(n,k)\in\{2,\ldots,6\}\times\{2,\ldots,5\}$ with $N_s{=}150$ instances per cell, $3{,}000$ in total, on the modal ROM of Use Case II. Each instance draws $k$ random ellipses, corridors, or half-planes, an initial state with $\htcap(\xi_0) > 0.015$, and $n$ waypoints inside $\bigcap_i\Safe_i$. Phases last $2$\,s, and the final phase lasts $4$\,s. We use $\alpha{=}2.0$, $\varepsilon{=}0.05$, and $\Delta t{=}0.02$\,s. A run is safe if $\min_i h_i \geq -10^{-3}$ throughout. We fit $p_0(k)$ by nonlinear least squares.

\begin{figure}[t]
  \centering
  \includegraphics[width=0.9\columnwidth]{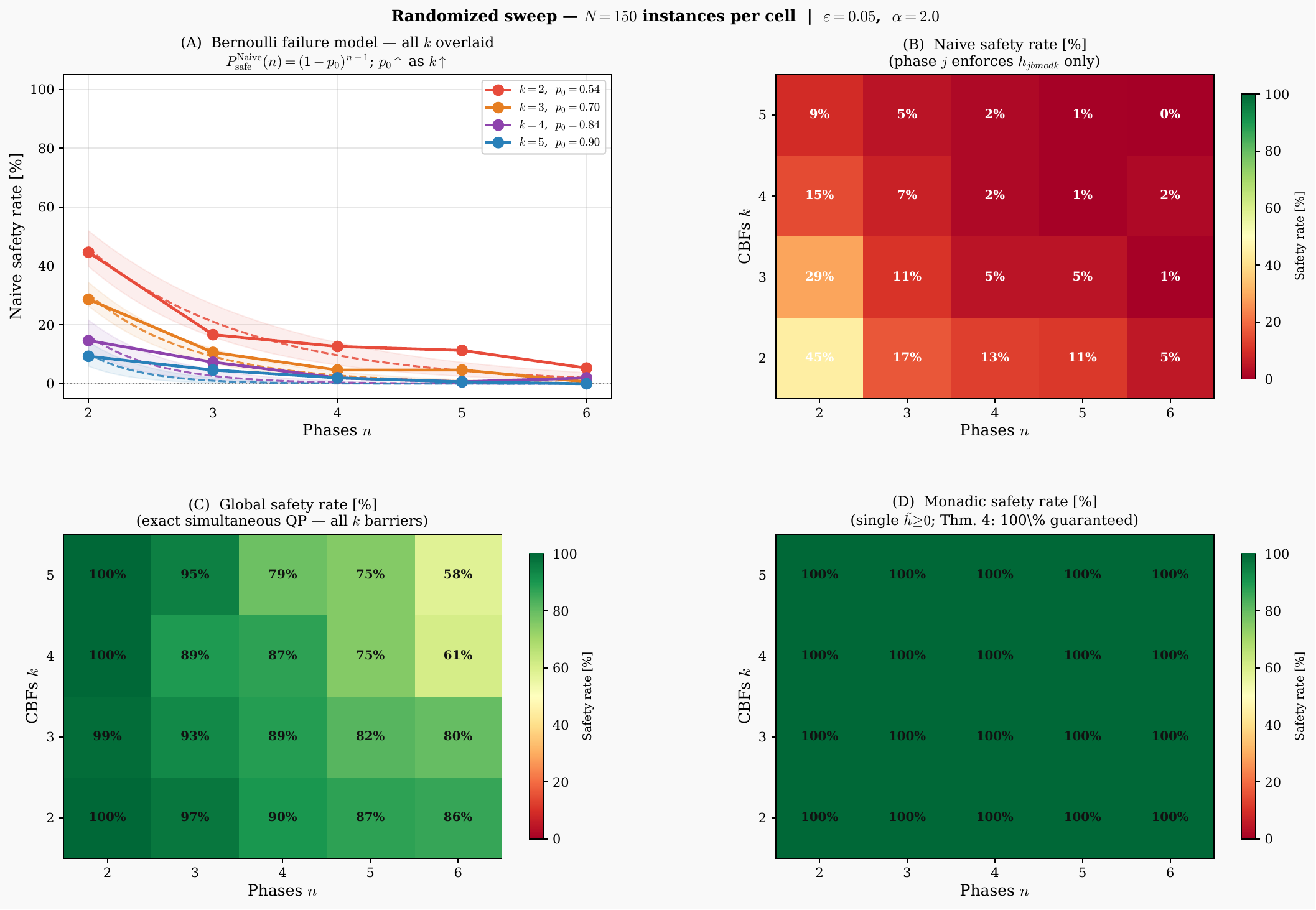}
  \caption{Randomized sweep over $N_s{=}150$ random CBF geometries per cell, with $\varepsilon{=}0.05$ and $\alpha{=}2.0$. (A)~Naive safety rate versus phase count $n$ for each barrier count $k$. Dashed curves show the fitted Bernoulli model $P_{\mathrm{safe}}^{\mathrm{Naive}}(n){=}(1-p_0)^{n-1}$, and shaded regions show $p_0 \pm 2$ standard errors. The fitted $p_0$ increases monotonically with $k$ (Remark~\ref{rem:p0geom}). (B)--(D)~Safety rate over the full $(n,k)$ grid for Naive, Global, and Monadic. Naive collapses exponentially with $n$. Global degrades under constraint conflict as $k$ grows, despite being the exact simultaneous-QP baseline. Monadic achieves $100\%$ on all 20 cells, as Theorem~\ref{thm:main} guarantees.}
  \label{fig:sweep}
\end{figure}

\textbf{Results.} Figure~\ref{fig:sweep} summarizes the sweep. The fitted values $p_0 = 0.54, 0.70, 0.84, 0.90$ for $k = 2, \dots, 5$ increase monotonically with $k$ (Remark~\ref{rem:p0geom}). Naive drops to single digits by $n{=}6$ for every $k$. Monadic achieves $100\%$ on every cell, as Theorem~\ref{thm:main} guarantees.

Panel~(C) shows a failure mode that our framework does not capture. Global degrades as $k$ and $n$ grow, reaching $58\%$ at $k{=}5$, $n{=}6$. When the $k$ constraint normals conflict in the two-dimensional input space, the QP becomes infeasible and Global falls back to the nominal input. This does not contradict Theorem~\ref{thm:main}. Sub-monad membership presupposes that a certified controller exists, and Global lands in $\TC{\bigcap_i \Safe_i}$ only while its QP is feasible. Input-constrained compatibility conditions~\cite{breeden2023compositions, tan2022compatibility} address exactly this existence question. Softmin composition does not guarantee Monadic feasibility in general~\cite{rabiee2023softmin}, yet Monadic remained safe in all $3{,}000$ instances. Building feasibility into the sub-monad condition itself is an open question.



\section{Use Cases}
\label{sec:usecases}

We have established the categorical account and quantified how reliably it avoids the failure mode of naive composition. We now demonstrate the three strategies of \S\ref{sec:statistical} on two concrete robotic systems.

Both use cases compare three composition strategies. Naive has each phase controller $f_i$ enforce only the single barrier $h_i$, placing controllers in different sub-monads (Proposition~\ref{prop:naive_fails}). Global~\cite{glotfelter2017nonsmooth} enforces all $k$ barriers as separate hard QP constraints at every step, giving a shared sub-monad $\TC{\bigcap_i \Safe_i}$ at the cost of $k$ constraints per step. Monadic uses a single $\htcap$ constraint, giving the shared sub-monad $\TC{\{\htcap \geq 0\}} \subseteq \TC{\bigcap_i \Safe_i}$ of Theorem~\ref{thm:main} with one constraint per step. All CBF-QPs are solved with Gurobi at 50\,Hz.

\subsection{Use Case I: Four-Mode PWA System}

\textbf{Setup.} We use three safe sets, the ellipse $h_1 = 1 - x_1^2/0.92^2 - x_2^2/0.78^2$, the corridor $h_2 = 0.72^2 - x_1^2$, and the half-plane $h_3 = x_2 + 0.36$. Four phases of $(3.0, 2.0, 2.5, 8.0)$\,s take the state from $x_0 = (-0.50, 0.55)^\top$ in mode 2 to the goal $(0.35, -0.20)^\top$ in mode 4 under a saturated proportional nominal law. We use $\varepsilon{=}0.04$, $\alpha{=}1.5$, forward Euler with $\Delta t{=}0.015$\,s, and $N_s{=}50$ runs with initial states perturbed by $\mathcal{U}([-0.06, 0.06]^2)$.

\textbf{Failure mechanism.} The phase-1 target $(-0.65, 0.25)^\top$ lies near the left corridor wall, where the mode-2 drift points left. Naive enforces only $h_1$ in phase 1, so the drift carries $x_1$ past $-0.72$ before the first handoff. A controller in $\TC{\Safe_1}$ thus need not lie in $\TC{\Safe_1 \cap \Safe_2}$.

\begin{figure}[t]
  \centering
  \includegraphics[width=0.85\columnwidth]{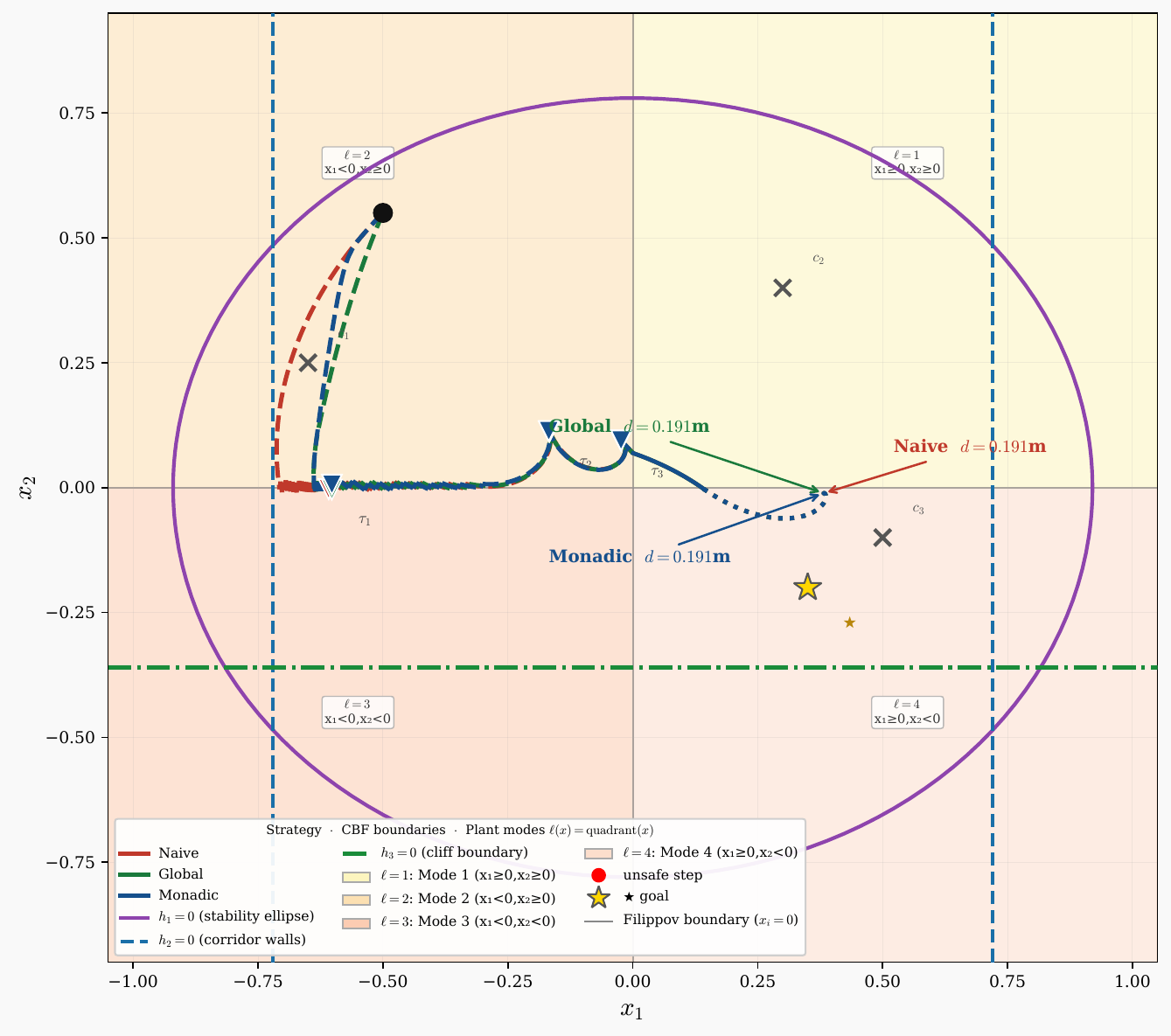}
  \caption{Four-mode PWA system, one representative rollout. Naive (red) exits $\Safe_1\cap\Safe_2\cap\Safe_3$ near the mode-2/mode-3 transition. Global (green) and Monadic (blue) remain safe throughout. Line style shows the plant mode $\ell(x)$, which switches as the trajectory crosses quadrants. Distances are for this single rollout, not the $N_s{=}50$ Monte Carlo averages.}
  \label{fig:pwa_traj}
\end{figure}

\textbf{Results.} Naive violates safety in 78\% of runs. Monadic matches the 100\% safety of Global with one constraint per step instead of three. It also keeps a larger margin, $\min h_1 = +0.09$ versus $+0.04$, consistent with $\delta = \varepsilon\ln 3 = 0.044$. Figure~\ref{fig:pwa_traj} shows a representative rollout.

\subsection{Use Case II: Unicycle ROM on Heterogeneous Terrain}

\textbf{Setup.} The ROM of Use Case II crosses a Rocky Terrain Traverse in four phases of $(2.0, 2.5, 2.5, 5.0)$\,s. It must respect the ellipsoidal margin $h_1 = 1 - p_x^2/1.20^2 - p_y^2/0.88^2$, the stepping-stone corridor $h_2 = 0.55^2 - p_x^2$, and the cliff $h_3 = p_y + 0.28$. A proportional nominal controller drives to phase targets inside $\bigcap_i\Safe_i$, at constant speed in phases 1 to 3. We use $\varepsilon{=}0.05$, $\alpha{=}2.0$, $\Delta t{=}0.02$\,s, and $N_s = 50$ runs.

\begin{figure}[t]
  \centering
  \includegraphics[width=0.85\columnwidth]{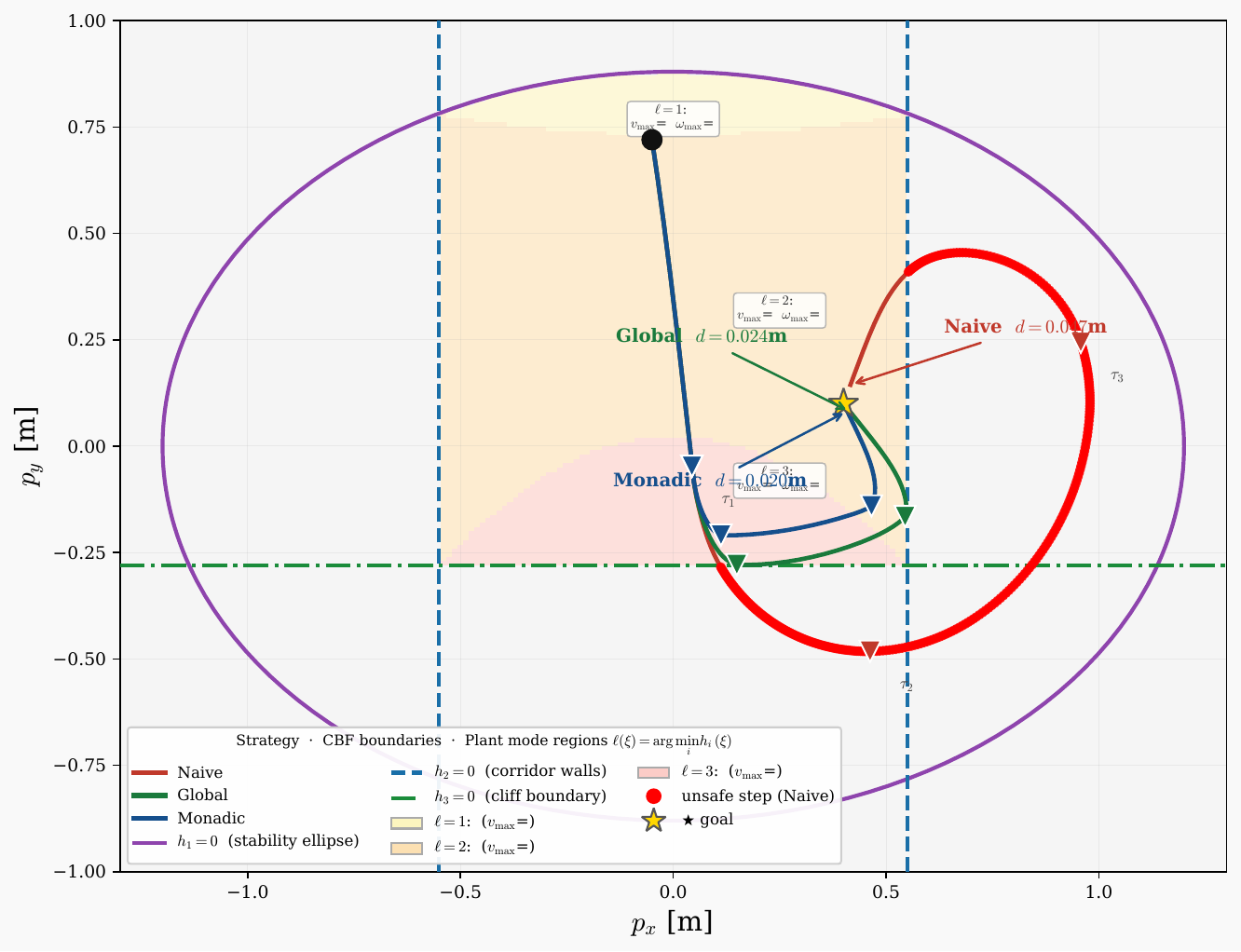}
  \caption{Rocky Terrain Traverse, one representative rollout. Naive (red) loops through mode $\ell{=}3$ near the handoffs $\tau_2$ and $\tau_3$ and crosses the cliff boundary $h_3{=}0$. Global (green) and Monadic (blue) remain within $\Safe_1\cap\Safe_2\cap\Safe_3$ and reach the goal. Distances are for this single rollout, not the $N_s{=}50$ Monte Carlo averages.}
  \label{fig:rocky_traverse}
\end{figure}

\textbf{Failure mechanism.} The heading $\psi$ carries across each handoff. Naive enforces $h_1$, $h_2$, $h_3$, and $h_1$ in turn, so nothing constrains the cliff in phases 2 and 4. Near $\tau_2$ and $\tau_3$, the inherited heading carries the robot across $h_3 = 0$ (Fig.~\ref{fig:rocky_traverse}). The composite lands in $\TT(M)$ but not in $\TC{\bigcap_i \Safe_i}$.

\textbf{Results.} Naive is never safe, with $\min h_3 = -0.343$. Monadic matches the 100\% safety of Global with one constraint per step. Its margins, $\min h_2 = +0.077$ and $\min h_3 = +0.070$, exceed $\delta = \varepsilon\ln 3 = 0.055$. Global's margins are only $+0.002$ and $+0.001$.

\textbf{Results.} Naive achieves 0\% safety, with $\min h_3 = -0.343$. This is a structural, not marginal, violation, confirming Proposition~\ref{prop:naive_fails}. Monadic matches Global safety at 100\% with one QP constraint per step. The Monadic minimum margins, $\min h_2 = +0.077$ and $\min h_3 = +0.070$, are both tightly consistent with the theoretical lower bound $2\varepsilon\ln 2 = 0.069$. This validates that the controller is conservative by \emph{exactly} the shift of Remark~\ref{rem:shift}, no more and no less. Global, by contrast, attains margins only marginally above zero, $+0.002$ and $+0.001$. Monadic therefore achieves the same safety guarantee with quantified, structural conservatism rather than incidental margin.

\section{Conclusion}
\label{sec:conclusion}
We constructed the trajectory monad $(\TT,\un,\mul)$ for Filippov-regular differential inclusions. We showed that CBF certificates lift to sub-monads, so controllers that land in a common sub-monad compose safely. Naive phase-by-phase composition fails because successive controllers land in different sub-monads. This mismatch is localized at each Kleisli handoff. It motivates a closed-form Bernoulli failure model, which we validated over $3{,}000$ randomized instances. A smooth intersection CBF with an explicit margin places all phase controllers in one shared sub-monad, with one QP constraint per step. It achieved $100\%$ safety in the sweep and on both use cases. The standard simultaneous-constraint baseline, in contrast, degraded under constraint conflict as the barrier count grew. Future work includes incorporating QP feasibility into the sub-monad condition, a sampled-data derivation of the margin, and Eilenberg--Moore algebras of $\TT$, which we conjecture to be Lyapunov certificates. We also plan hardware validation on a Unitree Go2 quadruped.

\bibliographystyle{IEEEtran}
\bibliography{references}

\end{document}